\documentclass[conference]{IEEEtran}

\pdfoutput=1 
\usepackage{hyperref}
\usepackage{color,overpic,cite}
\usepackage{pdfsync}
\usepackage{url}
\usepackage{amsmath,amsfonts,amssymb,amsthm,bbm,bm,comment,multirow,subfigure}
\usepackage{algorithm}
\usepackage{algorithmic}
\usepackage{tcolorbox}

\newcommand{\eat}[1]{{}}
\newcommand{\mean}[1]{\mathbb{E}\!\left[#1\right]}
\newcommand{\prob}[1]{\mathbb{P}\!\left(#1\right)}
\newcommand{\argmin}{\arg\min}

\newtheorem{theorem}{Theorem}
\newtheorem{lemma}{Lemma}
\newtheorem{proposition}{Proposition}

\newtheorem{definition}{Definition}

\newtheorem{corollary}{Corollary}

\newcommand{\ev}{{\bf e}}

\newcommand{\Yc}{{\cal Y}}
\newcommand{\Xc}{{\cal X}}
\newcommand{\Sc}{{\cal S}}
\newcommand{\Nc}{{\cal N}}

\newcommand{\Mc}{{\cal M}}

\newcommand{\Sigmam}{\hbox{\boldmath$\Sigma$}}

\usepackage{enumitem}

\title{Learning to Cache With No Regrets} 

\vspace{-3mm}
\author{\IEEEauthorblockN{Georgios S. Paschos\IEEEauthorrefmark{1}, Apostolos Destounis\IEEEauthorrefmark{1}, Luigi Vigneri\IEEEauthorrefmark{1}\IEEEauthorrefmark{2}, George Iosifidis\IEEEauthorrefmark{3}}\\
	\vspace{-4.5mm}
	\IEEEauthorblockA{
		\IEEEauthorrefmark{1}France Research Center, Huawei Technologies, Paris\\
		\IEEEauthorrefmark{2}IOTA Foundation, Berlin, Germany\\
		\IEEEauthorrefmark{3}School of Computer Science and Statistics, Trinity College Dublin, Ireland}
	\thanks{{This work was supported by Science Foundation Ireland under Grant No. 17/CDA/4760. The opinions expressed in this paper are of the authors alone, and do not represent an official position of Huawei Technologies. }}%
	\vspace{-0.5em}
}

\IEEEoverridecommandlockouts
\begin{document}
\maketitle
\begin{abstract}
This paper introduces a novel caching analysis that, contrary to prior work, makes no modeling assumptions for the file request sequence. We  cast the caching problem in the framework of \emph{Online Linear Optimization} (OLO), and introduce a class of \emph{minimum regret} caching policies, which minimize the losses with respect to the \emph{best static configuration in hindsight} when the request model is unknown. These policies are very important since they are robust to popularity deviations in the sense that they learn to adjust their caching decisions when the popularity model changes. We first prove a novel lower bound for the regret of any caching policy, improving existing OLO bounds for our setting. Then we show that the  \emph{Online Gradient Ascent} (OGA) policy guarantees a regret that matches the lower bound, hence it is universally optimal. Finally, we shift our attention to a network of caches arranged to form a bipartite graph, and show that the  \emph{Bipartite Subgradient Algorithm} (BSA)  has no regret.
\end{abstract}

\section{Introduction}

We study the performance of caching systems from a new perspective: \emph{we seek a caching policy that optimizes the system's performance under any distribution of file request sequence}. This not only has huge practical significance as it tackles the caching policy design problem in its most general form, but also reveals a novel connection between caching and online linear optimization \cite{zinkevich2003online,Shalev12,Mert18}. This, in turn, paves the way for a new mathematical framework enabling the principled design of caching policies with performance guarantees.

\subsection{Background and Related Work}

Due to its finite capacity a cache can typically host only a small subset of the file library, and hence a \emph{caching policy} must decide which files should be stored. The main performance criterion for a caching policy is the so-called \emph{cache hit ratio}, i.e., the portion of file requests the cache can satisfy locally. Several policies have been proposed in the past  with the aim to maximize the cache hit ratio. For instance, the Least-Recently-Used (LRU) policy inserts in the cache the newly requested file and evicts the one that has not been requested for the longest time period; while the Least-Frequently-Used (LFU) policy evicts the file that is least frequently requested. These policies (and variants) were designed empirically, and one might ask:  \emph{under what conditions do they achieve high hit ratios}? 

The answer depends on the properties  of the file request sequence. For instance, we know that \emph{(i)} for stationary requests, LFU achieves the highest hit ratio \cite{Fricker12}; \emph{(ii)} a more sophisticated \emph{age-based-threshold} policy maximizes the hit ratio when the requests follow the Poisson Shot Noise model \cite{snm}; and \emph{(iii)} LRU has the highest competitive hit ratio \cite{Sleator85} for the adversarial model \cite{Belady66, Mattson70} which assumes that the requests are set by an \emph{adversary} aiming to degrade the system's performance. However, the  highest competitive hit ratio is achieved by any  \emph{marking} policy \cite{Borodin98} -- including FIFO -- suggesting  that this metric is  perhaps ``too strong''  to allow a good policy classification. Evidently, to decide which policy  to use, it is necessary to know the underlying file request model,  which in practice is \emph{a priori} unknown and, possibly, time-varying. \emph{This renders imperative the design of a {universal} caching policy that will work provably well for all request models. } 

\begin{figure}
	\centering
	\includegraphics[width=3.3in]{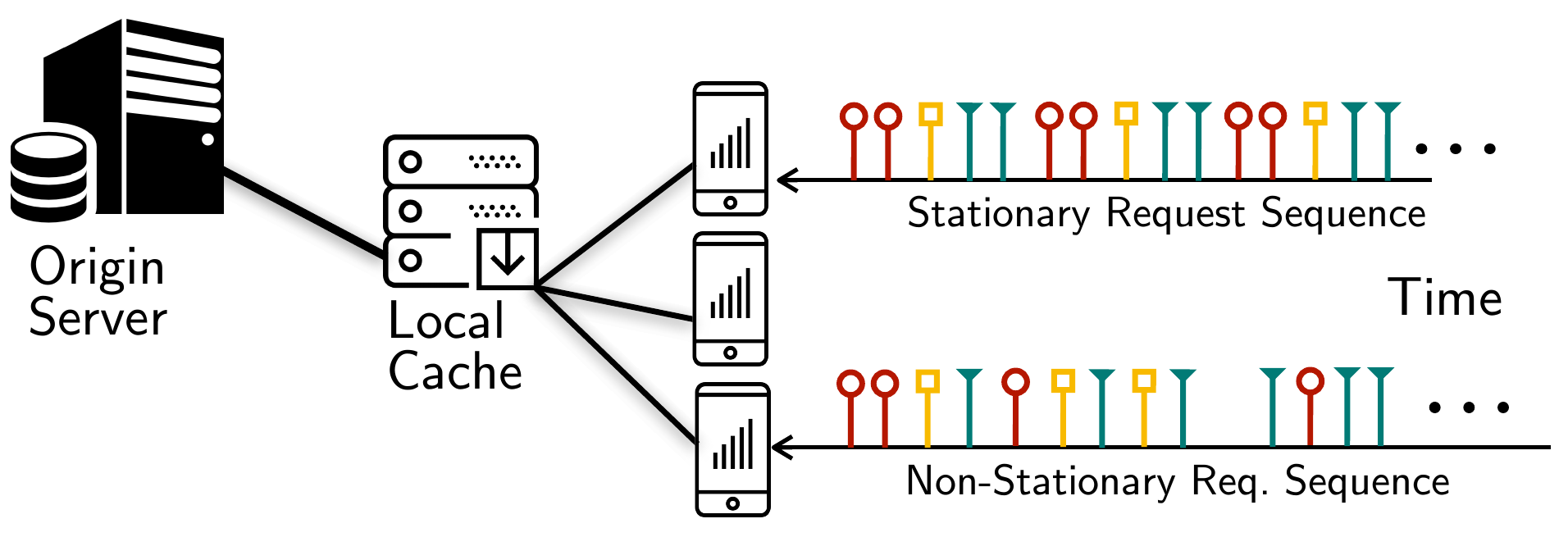}
	\vspace{-0.12in}\caption{File requests must  be served by  a local cache (hit) or the origin server (miss). A caching policy observes past requests and decides which files to cache in order to increase the hit ratio. \vspace{-0.15in}
	}
	\label{Fig:motivation}
\end{figure}

This is even more important in the emerging edge caching architectures that use small local servers or caches attached to wireless stations. These caches receive a low number of requests and therefore, inevitably, ``see'' request processes with highly non-stationary popularity \cite{snm, mathieu, Elayoubi2015}. Prior works employ {random replacement models \cite{Elayoubi2015} or inhomogeneous Poisson processes \cite{snm,kauffman} to model the requests in these systems. However, such multi-parametric models are challenging to fit and rely on strong assumptions about the popularity evolution}. Other notable  approaches learn the  instantaneous popularity model with no prior assumptions. For instance \cite{gunduz-reinforcement,giannakis-q-learning} employ \emph{Q-learning}, and \cite{mihaela-video-caching} leverages a scalable prediction method; but contrary to our approach, they assume the popularity evolution is stationary across time. The design of adaptive paging policies is systematically studied  in \cite{geulen2010regret,englert2013economical, lykouris-ML} as an online learning problem. These works, however, consider only the basic paging problem, i.e., hit ratio maximization when caching entire files in a single cache.

Caching networks (CNs) on the other hand, are hitherto under-explored; yet very important as most often caches are organized in networks. In CNs one needs to jointly decide which cache will satisfy a request (routing) and which files will be evicted (caching). The works \cite{ioannidis_icn17, Poularakis2014Approximation} proposed joint routing and caching policies for bipartite and general network graphs, respectively, and \cite{jungho-wiopt} extended them to CNs with elastic storage. However all these works assume that file popularity is stationary. On the other hand, \cite{giovanidis-mLRU} proposed the multi-LRU (mLRU) strategy, and \cite{leonardi-implicit} proposed the ``lazy rule'' extending $q$-LRU to provide local optimality guarantees under stationary requests. \emph{It transpires that we currently lack a principled method to design caching policies with provable performance guarantees, under general popularity conditions, for single or networks of caches. This is exactly our focus.}


\vspace{-0.5mm}
\subsection{Methodology and Contributions}

We introduce a model-free caching policy along the lines of Online Linear Optimization (OLO). For the single cache case we obtain matching lower and upper regret bounds, proving that the online gradient ascent achieves order optimal performance, and we then extend these results to CNs. We assume that file requests are drawn from a general distribution, which is equivalent to caching versus an adversary selecting the requests. At each slot, in order, \emph{(i)} the caching policy decides which file parts to cache; \emph{(ii)} the adversary introduces the new request; and \emph{(iii)} a file-dependent utility is obtained proportional to the fraction of the request that was served by the cache. This generalizes the criterion of cache hit ratio and allows us to build policies that, e.g., optimize delay or provide preferential treatment to files. In this setting, we seek to find a caching policy with minimum \emph{regret}, {i.e., minimum utility loss over an horizon $T$, compared to the best cache configuration when the request sample path is known}.

We {prove} that well-known caching policies such as LRU and LFU have $\Omega (T)$ regret, hence there exist request patterns for which these policies fail to learn a good caching configuration (losses increase with time). In contrast, we propose the \emph{Online Gradient Ascent} (OGA) policy, and  prove its regret is at most $O(\sqrt{CT})$, for a cache that can store $C$ out of the $N$ total files. This shows that OGA eventually (as $T\to\infty$) amortizes the losses under any request  model, even under denial-of-service attacks. Additionally we prove a \emph{novel lower bound} which is tighter than the general OLO lower bound. 
 For the case of online hit ratio maximization, we find that any policy must have at least $\Omega(\sqrt{CT})$ regret.
  Combining the two results we conclude that 
\emph{(i)} the regret of online caching is exactly $\Theta(\sqrt{CT})$, and \emph{(ii)} OGA is a universally optimal policy. Interestingly, OGA can be seen as a regularized LFU, or a slightly modified LRU as we explain in Section \ref{sec:FoLreg}. 

We extend our model to a network of caches, arranged in the form of a bipartite graph -- a setting known as femtocaching \cite{golrezaei2012femtocaching}. We provide the \emph{Bipartite Subgradient Algorithm} (BSA) caching strategy that achieves a regret $O(\sqrt{\text{deg }JCT})$, where $J$ is the number of caches and $\text{deg}$ the maximum network node degree. Our contributions can be summarized as follows:
\begin{itemize}[leftmargin=3.5mm]
	\item \emph{Machine Learning (ML) approach}: we provide a fresh ML angle into caching policy design and performance analysis. To the best of our knowledge this is the first time OLO is used to provide optimality bounds in caching problem. Moreover, we reverse-engineer the standard caching LRU/LFU-type of policies, by drawing connections with OLO, and provide directions for improving them.
	
	
	\item  \emph{Universal single-cache policy}: The proposed OGA policy is universally optimal, i.e., provides zero loss versus the best caching configuration under any request model. An important projection algorithm is provided to reduce complexity and enable operation in large caches. 
	
	\item \emph{Universal bipartite caching}: We consider a general bipartite CN and design the \emph{online joint caching and routing BSA policy}. Our approach hits the sweet spot of complexity versus utility for CNs: offers rigorous performance results, while it is applicable to fairly complicated settings.
	
	\item \emph{Trace-driven Evaluation}: We employ a battery of tests evaluating our policies with several request patterns. We find that OGA outperforms LRU/LFU by 20$\%$ in different scenarios, while BSA beats lazy-LRU by 45.8$\%$.
\end{itemize}



\vspace{1mm}
\section{System model}\label{sec:model}
We study a system with a library of files $\mathcal{N}\!=\!\{1,2,\dots, N\}$ of equal size and a cache that fits $C<N$ of them. The system evolves in time slots, and in each slot $t$ a single request is made for $n\!\in\!\mathcal{N}$, denoted with the event $x_{t}^{n}\!=\!1$. Vector $x_t\!=\!(x_{t}^n, n\!\in\!\mathcal{N})$ represents the $t$-slot request, chosen from the set:
\[
\mathcal{X}=\left\{x\in \{0,1\}^N ~\bigg |~ \sum_{n=1}^N x^n=1\right\}.
\]
The instantaneous file popularity is determined by the probability distribution $P(x_t)$ (with support $\Xc$), which is allowed to be unknown and arbitrary; and the same holds for the joint distribution $P(x_1,\dots,x_T)$ that describes the file popularity evolution. This generic model captures all studied request sequences in the literature, including stationary (i.i.d. or otherwise), non-stationary, and adversarial models. 

The cache is managed with the caching configuration variable $y_t\in [0,1]^N$, that denotes the fraction of file $n$ cached in slot $t$.\footnote{Why caching of file fractions makes sense? Large video files are composed of chunks stored independently, see literature of \emph{partial caching} \cite{maggi}. Also, the fractional variables may represent caching probabilities \cite{Shalev12,Blaszczyszyn2014Geographic}, or coded equations of chunks \cite{golrezaei2012femtocaching}. For practical systems, the fractional $y_t^n$ should be rounded, which will induce a small application-specific error.} Taking into account the cache size $C$, the set $\mathcal{Y}$ of admissible caching  configurations is:
\begin{equation}\label{eq:capped}
\mathcal{Y}=\left\{y\in [0,1]^N ~\bigg |~ \sum_{n=1}^Ny^n\leq C\right\}.
\end{equation}

\begin{definition}[Caching Policy]
A caching policy $\sigma$ is a (possibly randomized) rule that at slot $t=1,\dots,T$ maps past observations $x_1,\dots,x_{t-1}$ and configurations $y_1,\dots,y_{t-1}$ to the configuration $y_t(\sigma)\in \mathcal{Y}$ of slot $t$.
\end{definition}

We denote with $w^n$ the  utility obtained  when file $n$ is requested and found in the cache (also called a \emph{hit}). This file-dependent utility can be used to model bandwidth economization from cache hits \cite{maggi}, QoS improvement \cite{golrezaei2012femtocaching}, or any other cache-related benefit. We will also be concerned with the special case $w^n=w, n\!\in\!\mathcal{N}$, i.e., the \emph{cache hit ratio} maximization. A cache configuration $y_t$ accrues in slot $t$ a utility $f\big(x_t,y_t\big)$ determined as follows:
\[
f\big(x_t,y_t\big)=\sum_{n=1}^N w^n x_t^ny^n_t.
\]

\begin{figure}[!t]
	\centering
	\includegraphics[width=3.5in]{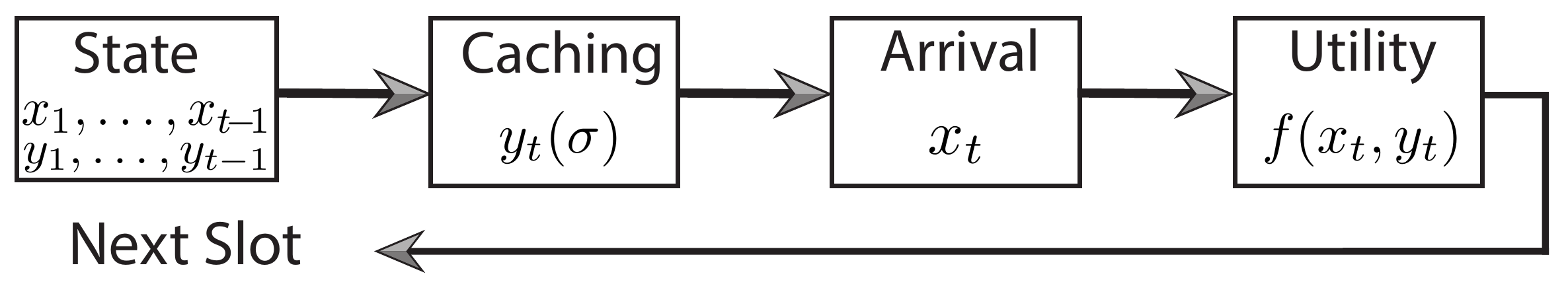}\vspace{-5mm}
	\caption{A caching decision is taken; the adversary selects the request; the caching utility is realized; and the system state is updated for next $t$.\vspace{-3mm}}
	\label{fig:model}
\end{figure}

Let us now cast caching as an online linear optimization problem. This requires the following conceptual innovations. Since we allow the request sequence to follow any arbitrary probability distribution, we may equivalently think of $x_t$ as being set by an \emph{adversary} that aims to degrade the performance of the policy. Going a step further, we can equivalently interpret that the adversary selects at each slot $t$ the utility function of the system $f_t(y)$ from a family of linear functions, $f_t(y)\equiv f(x_t,y)$. Finally, note that $y_t$ is set at the beginning of slot $t$, before the adversary selects $x_t$, Fig.~\ref{fig:model}. The above place our problem squarely in the OLO framework.

Given the adversarial nature of our model, the ability to extract useful conclusions depends crucially on the selected performance metric. Differently from the competitive ratio approach of \cite{Sleator85}, we introduce a new metric that compares how our caching policy fares against the \emph{best static policy in hindsight}. This metric is often used in the literature of machine learning \cite{Shalev12,Mert18} with the name {worst-case static regret}. In particular, we define the \emph{regret of policy $\sigma$} as:
\begin{align*}
R_T(\sigma)&=\max_{P(x_1,\dots,x_T)}
\mean{\sum_{t=1}^T f_t\big(y^*\big)-\sum_{t=1}^T f_t\big(y_t(\sigma)\big)} 
\end{align*}
where $T$ is the horizon; the maximization is over the admissible adversary distributions; the expectation is taken w.r.t. the possibly randomized $x_t$ and $y_t(\sigma)$; and $y^*\in\arg\max_{y\in \Yc}\sum_{t=1}^T f_t(y)$ is the best static configuration in hindsight, i.e., a benchmark policy that knows the sample path $x_1,\dots,x_T$. Intuitively, measuring the utility loss of $\sigma$ over $y^*$ constrains the power of the adversary: radical request pattern changes will challenge $\sigma$ but also induce huge losses in $y^*$. This comparison allows us to discern policies that can learn good caching configurations from those that fail. 

We seek a caching policy that minimizes the regret by solving the problem $\inf_{\sigma}R_T(\sigma)$, known as \emph{Online Linear Optimization} (OLO) \cite{Shalev12}. The analysis in OLO aims to discover how the regret scales with horizon $T$. A caching policy with sublinear regret $o(T)$ produces average losses $R_T(\sigma)/T\!\to\! 0$ with respect to the best configuration with hindsight, hence it learns to adapt the cache configuration without any prior knowledge about the request distribution. Our problem is further compounded due to the cache size dimension. Namely, apart from optimizing the regret w.r.t. $T$, it is of practical importance to consider also the dependence on $N$ (or $C$).  


\textbf{Regret of Standard Policies}. Having introduced this new caching optimization formulation, it is interesting to characterize the worst case performance of LRU and LFU policies. Recall that LRU caches the $C$ most recently requested files, while LFU calculates for each file the request frequency $h^n_t=\frac{1}{t-\tau^n}\sum_{i=\tau^n}^tx_i^n$,
where $\tau^n$ is the slot when file $n$ was requested for the  first time, and caches the $C$ most frequent files. The following proposition describes their performance under arbitrary requests where, for simplicity, we assumed $w^n=w,\forall n$ (hit ratio  maximization).
\begin{tcolorbox}[boxrule=0.7pt,arc=0.6em, left=1.5mm, right=1.5mm] 
\begin{proposition} 
The regret of \textup{LRU}, \textup{LFU}  satisfies:
\begin{align*}
&R_T(\textup{LRU})=R_T(\textup{LFU})\geq wC\left(\frac{T}{C+1}-1\right)\,.
\end{align*}
\end{proposition}
\end{tcolorbox}

\begin{IEEEproof} 
Assume that the adversary chooses the periodic request sequence $\{1,2,\dots,C+1,1,2,\dots\}$. For any $t>C$, since  the requested file is the $C+1$ least recent file, it is not in the LRU cache, and no utility is received. Hence, LRU can achieve at most $wC$ utility from the first $C$ slots. However, a static policy with hindsight  achieves at least $wTC/(C+1)$ by caching the first $C$ files. The same rationale can be used for LFU by noticing that due to the structure of the periodic arrivals, the least frequent file is also the least recent one.
\end{IEEEproof}
The $\Omega(T)$ performance of standard caching policies is poor, yet this is rather expected since they are designed to perform well only under certain models (LRU for requests with temporal locality; LFU for stationary requests), and they are known to under-perform in other cases, e.g., LRU in stationary, and LFU in decreasing popularity patterns. Low regret performance means that there exist request distributions for which the  policy fails to ``learn'' what is the best configuration. \emph{Remarkably, we show below that there exist universal caching policies which ensure low regret under any request model.}

\vspace{2mm}
\section{Regret Lower Bound}\label{sec:regret-bound}

A regret lower bound is a powerful theoretical result that provides the fundamental limits of how fast any algorithm can learn to cache, much like the upper bound of the channel capacity. Regret lower bounds in OLO have been previously derived for different action sets: for $N$-dimensional unit ball centered at the origin in \cite{Abernethy08}, and  $N$-dimensional hypercube in \cite{Hazan06}. In our case, however, the above results do not apply since  $\Yc$  in \eqref{eq:capped} is a capped simplex, i.e., the intersection of a box and a simplex inequality. Therefore, we need the following new regret lower bound tailored to the online caching problem. 

\begin{tcolorbox}[boxrule=0.7pt,arc=0.6em, left=1mm, right=1mm] 
\begin{theorem}[Regret Lower Bound]\label{th:lowerBoundRegret}
The regret of any  caching policy $\sigma$   satisfies: \vspace{-0.1in}
	\[
R_T(\sigma) > \sum_{i=1}^C\mean{Z_{(i)}}\sqrt{T},\quad \text{ as } T\to\infty, \vspace{-0.051in}
	\]
where $Z_{(i)}$ is the $i$-th maximum element of a Gaussian random vector with zero mean and covariance matrix $\Sigmam(w)$ given by \eqref{eq:covarianceMatrix}.  
	
Furthermore, assume $C<N/2$ and define $\phi$ any permutation of $\mathcal{N}$ and $\Phi$ the set of all such permutations:
\[
R_T(\sigma)>\frac{\max_{\phi\in\Phi}\sum_{k=1}^C\sqrt{w^{\phi(2(k-1) +1)}+ w^{\phi(2k)}}}{\sqrt{2\pi\sum_{n=1}^N1/w^n}}\sqrt{T}	
\] 
\end{theorem}
\end{tcolorbox}

In the important special case of hit rate maximization, where each file $n$ is $w^n\!=\!w$, the above bound simplifies to:
\begin{tcolorbox}[boxrule=0.7pt,arc=0.6em, colback=gray!3] 
\begin{corollary}\label{cor:lbound1}
Fix $\gamma\triangleq C/N$,  $w^n=w,~\forall n$, and $C<N/2$. Then, the regret of any caching policy $\sigma$  satisfies: 
	\[
		R_T(\sigma) > w\sqrt{\frac{\gamma}{\pi}}\sqrt{CT}, \quad \text{ as } T\to\infty.
	\]
\end{corollary}
\end{tcolorbox}

Before providing the proof, a few remarks are in order. Our bound is tighter than  the classical $\Omega(\sqrt{(\log N) T})$ of OLO in the literature  \cite{Hazan06,Abernethy08}, which is attributed to the difference of  sets $\Xc,\Yc$. In our proof we provide technical lemmas that are also useful, beyond caching, for the regret analysis of capped simplex sets. In next section we will design a caching policy that achieves regret  $O(\sqrt{CT})$, establishing that the regret of online caching is in fact $\Theta(\sqrt{CT})$.

\begin{IEEEproof}[Proof of Theorem \ref{th:lowerBoundRegret}]
To find a lower bound, we will analyze a specific adversary $x_t$. 
In particular, we will  consider an i.i.d.  $x_t$ 
such that file $n$ is requested with probability 
\[
\prob{x_t=\ev_n}=\frac{1/w^n}{\sum_{i=1}^N1/w^i}, ~~\forall n,t,
\] 
where $\ev_n$ is a vector with element $n$ equal to $1$ and the rest zero.
With such a choice of $x_t$, any causal caching policy yields an expected utility  at most $CT/\sum_{n=1}^N(1/w^n)$, since 
\begin{align}
\mean{\sum_{t=1}^T f_t(y_t(\sigma))}
&=\sum_{t=1}^T\sum_{n=1}^Nw^n\prob{x_t=\ev_n}y_t^n(\sigma)\label{eq:util}\\
&\hspace{-0.2in}=\sum_{t=1}^T\frac{1}{\sum_n 1/w^n}\sum_{n=1}^Ny_t^n(\sigma) 
\leq\frac{CT}{\sum_n 1/w^n}, \notag
\end{align}
independently of $\sigma$. 
To obtain a regret lower bound
we show that a static policy with hindsight can exploit the knowledge of the sample path $x_1,\dots,x_T$ to  achieve a higher utility than \eqref{eq:util}. Specifically, defining $\nu^n_t$ the number of times file $n$ is requested in slots $1,\dots,t$, the best static policy will cache the $C$ files with highest products $w^n\nu^n_T$. In the following, we characterize how this compares against the average utility of \eqref{eq:util} by analyzing the order statistics of an Gaussian vector.

{
For i.i.d. $x_t$ we may rewrite regret as the expected difference between the best static policy in hindsight and  \eqref{eq:util}: 
\begin{equation}\label{eq:regretIID}
R_T =\mean{ \max_{y\in\Yc}y^T\sum_{t=1}^T w \odot x_t }- \frac{CT}{\sum_n 1/w^n}
,\end{equation}
where $w \odot x_t=[w^1x^1_t, w^2x^2_t,...,w^Nx^N_t]^T$ is the Hadamard product between the weights and request vector. Further, \eqref{eq:regretIID} can be rewritten as a function:
\[
R_T=\mean{g_{N,C}(\overline{z}_T)} = \mean{\max_{b\in \stackrel{\circ}{\Yc}}\left[b^T\overline{z}_T\right]},
\]
where, \emph{(i)} $\stackrel{\circ}{\Yc}$ is  the set of all possible integer caching configurations (and therefore $g_{N,C}(.)$ is the sum of the maximum $C$ elements of its argument); and \emph{(ii)} the process $\overline{z}_T$ is  the vector of utility obtained by each file after the first $T$ rounds, centered around its mean:
\begin{align} \nonumber
\overline{z}_T & = \sum_{t=1}^{T}w \odot x_t - w \odot\frac{T}{\sum_{n=1}^N1/w^n}w^{-1} \\ \label{eq:centeredDemandVector}
& = \sum_{t=1}^{T}\left(z_t - \frac{1}{\sum_{n=1}^N1/w^n}\mathbf{1}_N\right) \end{align}
where $z_t = w \odot x_t$ are i.i.d. random vectors,  with distribution 
\[ \mathbb{P}\left(z_t = w^i \ev_i\right) = \frac{1/w^i}{\sum_{n=1}^N1/w^n}, \forall t, \forall i,
\]
and, therefore, mean $\mean{z_t}=\frac{1}{\sum_{n=1}^N1/w^n}\mathbf{1}_N$.\footnote{Above we have used the notation $w^{-1} = \left[1/w^1, 1/w^2,...,1/w^N\right]^T.$} 
}
A key ingredient in our proof is the limiting behavior of $g_{N,C}(\overline{z}_T)$:
\begin{lemma}\label{lem:regretConvDistribution}
	Let ${Z}$ be a Gaussian vector  $\Nc\left(\mathbf{0}, {\Sigmam}(w)\right)$, where ${\Sigmam}(w)$ is given in \eqref{eq:covarianceMatrix}, and ${Z}_{(i)}$ its $i-$th largest element. Then
	\[
\frac{g_{N,C}(\overline{z}_T)}{\sqrt{T}} \xrightarrow[T\rightarrow\infty]{\text{distr.}} \sum_{i=1}^C{Z}_{(i)}
	.\]
\end{lemma}  
\begin{IEEEproof}
Observe that $\overline{z}_T$ is the sum of $T$ uniform i.i.d. zero-mean random vectors, where the covariance matrix  can be calculated using \eqref{eq:centeredDemandVector}: ${\Sigmam}(w) =$
\begin{align}\nonumber
&=\mean{\left(z_1 - \frac{1}{\sum_{n=1}^N1/w^n}\mathbf{1}_N\right)\left(z_1 - \frac{1}{\sum_{n=1}^N1/w^n}\mathbf{1}_N\right)^T}\\ \label{eq:covarianceMatrix}
&= \frac{1}{\sum_{n=1}^N1/w^n}\begin{cases}
w_i -\frac{1}{\sum_{n=1}^N1/w^n}, i=j\\
-\frac{1}{\sum_{n=1}^N1/w^n}, i\neq j
\end{cases}
,\end{align}
where the second equality follows from the distribution of  $z_t$ and some calculations.\footnote{For the benefit of the reader, we note that $Z$ has no well-defined density (since ${\Sigmam}(w)$ is singular). For the proof, we only use its distribution.}
Due to the Central Limit Theorem: 
\begin{equation}\label{eq:convCenteredX}
\frac{\overline{z}_T}{\sqrt{T}} \xrightarrow[T\rightarrow\infty]{\text{distr.}} {Z}.\end{equation}
Since $g_{N,C}(x)$ is continuous, \eqref{eq:convCenteredX}  and the Continuous Mapping Theorem \cite{Billingsley99} imply
\[
\frac{g_{N,C}\left({\overline{z}_T}\right)}{\sqrt{T}} = g_{N,C}\left(\frac{\overline{z}_T}{\sqrt{T}}\right) \xrightarrow[T\rightarrow\infty]{\text{distr.}} g_{N,C}\left({Z}\right)
,\] 
and the proof is completed by noticing that $g_{N,C}(x)$ is the sum of the maximum $C$ elements of its argument. 
\end{IEEEproof}
An immediate consequence of Lemma \ref{lem:regretConvDistribution},  is that 
\begin{equation}\label{eq:convExpectations}
\frac{R_T}{\sqrt{T}}=\frac{\mean{g_{N,C}(\overline{z}_T)}}{\sqrt{T}}\xrightarrow{T\rightarrow\infty}\mean{\sum_{i=1}^C{Z}_{(i)}} = \sum_{i=1}^{C}\mean{{Z}_{(i)}}
\end{equation}
and the first part of the Theorem is proved. 

To prove the second part, we remark that the RHS of \eqref{eq:convExpectations} is the expected sum of $C$ maximal elements of vector $Z$, and hence larger than the expected sum of any $C$ elements of $Z$. In particular, we will compare with the following: Fix a permutation $\bar{\phi}$ over all $N$ elements, partition the first $2C$ elements in pairs by combining 1-st with 2-nd, ..., $i$-th with $i$+1-th, $2C$-1-th with $2C$-th, and then from each pair choose the maximum element and return the sum. We then obtain: 
\begin{align}\nonumber
\mean{\sum_{i=1}^C{Z}_{(i)}} &\geq \mean{\sum_{i=1}^C\max\left[Z^{\bar{\phi}(2(i-1)+1)}, Z^{\bar{\phi}(2i)}\right]} \\ \nonumber
& = \sum_{i=1}^C\mean{\max\left[Z^{\bar{\phi}((2(i-1)+1)}, Z^{\bar{\phi}(2i)}\right]}
,\end{align}
where the second step follows from the linearity of the expectation, and the expectation is taken over the marginal distribution of a vector with two elements of $Z$. We now focus on  $\max\left[Z^{k}, Z^{\ell}\right]$ for (any) two fixed $k,\ell$. We have that 
\[
(Z^{k}, Z^{\ell})^T \sim \Nc\left(\mathbf{0}, \Sigmam(w_k, w_{\ell})\right)
\]  
where \vspace{-0.25in}
\begin{align*}
&\hspace{0.3in}\Sigmam(w^k, w^{\ell})= \\&= \frac{1}{\sum_{n=1}^N1/w^n}\begin{bmatrix}
w^k-\frac{1}{\sum_{n=1}^N1/w^n} & -\frac{1}{\sum_{n=1}^N1/w^n}\\
-\frac{1}{\sum_{n=1}^N1/w^n} & w^{\ell}-\frac{1}{\sum_{n=1}^N1/w^n}
\end{bmatrix}
.\end{align*}
From \cite{Clark1961} we then have:
\[
\mean{\max\left[Z^{k}, Z^{\ell}\right]} = \sqrt{\frac{1}{\sum_{n=1}^N1/w^n}}\frac{1}{\sqrt{2\pi}}\sqrt{w^{k}+w^{\ell}}
,\]
therefore: 
\begin{equation}\label{eq:orderStatTransformed}
\mean{\sum_{i=1}^C{Z}_{(i)}} \geq \frac{1}{\sqrt{2\pi}}\frac{\sum_{i=1}^C\sqrt{w^{\bar{\phi}((2(i-1)+1)}+w^{\bar{\phi}(2i)}}}{\sqrt{\sum_{n=1}^N1/w^n}},
\end{equation}
for all $\bar{\phi}$. 
The result follows noticing that the  tightest bound is obtained by maximizing   \eqref{eq:orderStatTransformed}  over all permutations. 
\end{IEEEproof}

\vspace{2mm}
\section{Online Gradient Ascent}\label{sec:gradient}

Gradient-based algorithms are widely used in resource allocation mechanisms due to their attractive scalability properties. Here, we focus on the online variant and show that, despite its simplicity, it achieves the best possible regret.

\subsection{Algorithm Design and Properties}

Recall that the utility in slot $t$ is described by the linear function $f_t(y_t)=\sum_{n=1}^Nw^nx_t^ny_t^n$. The gradient $\nabla f_t$ at $y_t$ is an $N$-dimensional vector with coordinates:
\[
\frac{\partial f_t}{\partial y_t^n}=w^n x^n_t,~n=1,\dots,N.
\]
\begin{definition}[OGA]
The \emph{Online Gradient Ascent} (OGA) caching policy adjusts the decisions ascending in the direction of the gradient:\vspace{-0.15in}
\[
y_{t+1}=\Pi_{\mathcal{Y}}\left(y_t+\eta_t \nabla f_t\right),
\]
where $\eta_t$ is the stepsize, and $\Pi_{\mathcal{Y}}\left(z\right)\triangleq \argmin_{y\in\Yc}\|z-y \|$ 
is the Euclidean projection of the argument vector $z$ onto $\mathcal{Y}$, and $\|.\|$ the Euclidean norm.
\end{definition}
The projection step is discussed in detail next. We emphasize that OGA bases the decision $y_{t+1}$ on the caching configuration $y_t$ and the most recent request $x_t$. Therefore, it is a very simple causal policy that does not require memory for storing the entire state (full history of $x$ and $y$). 

Let us now discuss the regret performance of OGA. We define first the set diameter $\textit{diam}(\mathcal{S})$  to be the largest Euclidean distance between any two elements of set $\mathcal{S}$. To determine the diameter, we inspect two vectors $y_1,y_2\in \Yc$ which cache entire and totally different files and obtain:
\[
\textit{diam}(\mathcal{Y})=\left\{
\begin{array}{ll}
\sqrt{2C} & \text{if } 0<C\leq N/2,\\
\sqrt{2(N-C)} & \text{if } N/2<C\leq N. 
\end{array}
\right.
\]
Also, let $L$ be an upper bound of $\|\nabla f_t\|$, we have $L\leq\max_n(\sum_n w^nx_t^n)\leq \max_n(w^n)\equiv w^{(1)}$.
\begin{tcolorbox}[boxrule=0.7pt,arc=0.6em] 
\begin{theorem}[Regret of OGA]\label{th:2}
Fix stepsize $\eta_t=\frac{\textit{diam}(\mathcal{Y})}{L\sqrt{T}}$,  the regret of \textup{OGA} satisfies:
\[
R_T(\textup{OGA})\leq \textit{diam}(\mathcal{Y})L{\sqrt{T}}.
\]
\end{theorem}
\end{tcolorbox}

\begin{IEEEproof}
Using the non-expansiveness property of the Euclidean projection \cite{Ber99book} we can bound the distance of the algorithm iteration from the best static policy in hindsight:
\begin{align*}
\|y^{t+1}\!-\!y^*\|^2&\triangleq\|\Pi_{\mathcal{Y}}\left(y_t\!+\!\eta_t \nabla f_t\right)\!-\!y^*\|^2\!\leq\! \|y_{t}\!+\!\eta_t\nabla f_t\!-\!y^*\|^2\notag\\
&= \|y_t-y^*\|^2+2\eta_t{\nabla f_t}^T(y_t-y^*)+\eta_t^2\|\nabla f_t\|^2, 
\end{align*}
where we expanded the norm. If we fix $\eta^t=\eta$ and sum telescopically over horizon $T$, we obtain:
\begin{equation*}
\|y_{T}-y^*\|^2\!\!\leq\! \|y_1-y^*\|^2+2\eta\sum_{t=1}^T{\nabla f_t}^T(y_t-y^*)+\eta^2\sum_{t=1}^T\!\|\nabla f_t\|^2.
\end{equation*}
Since $\|y_{T}-y^*\|^2\geq 0$, rearranging terms and using $\|y_1-y^*\|\leq \textit{diam}(\Yc)$ and $\|\nabla f_t\|\leq L$: 
\begin{equation}\label{eq:teleonl}
\sum_{t=1}^T{\nabla f_t}^T(y^*-y_t)\leq \frac{\textit{diam}(\Yc)^2}{2\eta} +\frac{\eta TL^2}2.
\end{equation}

For  $f_t$ convex it holds $f_t(y_t)\geq f_t(y)+ {\nabla f_t}^T(y_t-y)$, $\forall y\in\mathcal{Y}$, and with equality if $f_t$ is linear. Plugging these in the OGA regret expression ($\max$ operator is removed) we get:
\begin{align*}
R_T(OGA)&=\sum_{t=1}^T(f_t(y^*)-f_t(y_t)) =\sum_{t=1}^T{\nabla f_t}^T(y_t-y^*)\\
&\stackrel{\eqref{eq:teleonl}}{\leq} \frac{\textit{diam}(\Yc)^2}{2\eta} +\frac{\eta TL^2}2,
\end{align*}
and for $\eta=\textit{diam}(\Yc)/L\sqrt{T}$ we obtain the result. 
\end{IEEEproof}
Using the above values of $L$ and $\textit{diam}(\mathcal{Y})$ we obtain:
\[
R_T(OGA)\leq {w^{(1)}\sqrt{2CT}},\,\,\,\,\text{for}\,\, C<N/2\,.
\]
\begin{tcolorbox}[boxrule=0.7pt,arc=0.6em] 
\begin{corollary}[Regret of Online Caching]\label{cor:2}
Fix $C/N=\gamma$, $w^n=w$, for all $n$, and assume $C<N/2$, the regret of online caching satisfies:
\[
w\sqrt{\frac{\gamma}{\pi}} \sqrt{CT}\leq \min_{\sigma}R_T(\sigma) \leq w\sqrt{2}\sqrt{CT}~~~\text{as}~T\to\infty.
\]
\end{corollary}
\end{tcolorbox}
\noindent Corollary \ref{cor:2} follows from  Corol. \ref{cor:lbound1} and Theorem \ref{th:2}. We conclude that disregarding $\sqrt{2\pi/\gamma}$ (amortized by $T$) OGA achieves the best possible regret and thus fastest learning rate.

\subsection{Projection Algorithm}

We explain next the Euclidean projection $\Pi_{\mathcal{Y}}$ used in  OGA, which can be written as a constrained quadratic program:
\begin{align}
\Pi_{\mathcal{Y}}\left(z\right)\triangleq \argmin_{y\geq \bm{0}} & \sum_{n=1}^N (z^n-y^n)^2 \label{eq:projection}\\
\text{s.t. }& \sum_{n=1}^N y^n \leq C~\text{ and}~y^n\leq 1,~\forall n\in\mathcal{N}.\notag 
\end{align}
In practice $N$ is expected to be large, and hence we require a fast algorithm. 
Let us introduce the Lagrangian:
\begin{align}
	L(y, \rho, \mu, \kappa )&=\sum_{n=1}^N(z^n-y^n)^2+\rho(\sum_{n=1}^Ny^n-C)\nonumber \\ &+\sum_{n=1}^N\mu_n(y^n-1) - \sum_{n=1}^N\kappa_ny_n,\nonumber
\end{align}
where $\rho,\mu,\kappa$ are the  Lagrangian multipliers. The KKT conditions of \eqref{eq:projection} ensure that  the values of $y^n$  at optimality will be partitioned into three sets $\Mc_1, \Mc_2, \Mc_3$:
\begin{align}
&\Mc_1=\{n\in\mathcal{N}: y^n\!=\!1  \},\,\,\, \Mc_2\!=\!\{n\in\mathcal{N}:\! y^n=z^n\!-\!\rho/2\},\nonumber\\
 &\Mc_3\!=\!\{n\in\mathcal{N}:\! y^n\!=\!0 \},\label{eq:proj2} 
\end{align}
where $\rho=2\big(|\Mc_1|-C+\sum_{n\in \Mc_2} z^n\big)/|\Mc_2|$ follows from the tightness of  the simplex constraint. It suffices for the projection to determine a partition of files into these sets. Note that given a candidate partition, we can check in linear time whether it satisfies all KKT conditions (and only the unique optimal partition does). Additionally, one can show that the ordering of files in $z$ is preserved at optimal $y$, hence a known approach is to search exhaustively over all possible ordered partitions, which takes $O(N^2)$ steps \cite{wang2015projection}. For our problem, we propose Algorithm 1, which exploits the property that all elements of $z$ satisfy $z^n\leq 1$ except at most one (hence also $|\mathcal{M}_1|\in \{0,1\}$), and computes the projection in $O(N\log N)$ steps (where the term $\log N$ comes from sorting $z$). In our simulations each loop is visited at most two times, and the OGA simulation  takes comparable time with LRU.

Finally, the projection algorithm provides insight into  OGA functionality. In a slot where file $n'$ has been requested, OGA will increase $y^{n'}$ according to the stepsize, and then decrease all other variables $y_n>0, n\neq n'$ symmetrically until the simplex constraint is satisfied.

\begin{algorithm}[t]
\caption{Projection on Capped Simplex} 
\label{alg1} 
\begin{algorithmic}[1] 
    \REQUIRE $C$; sorted $z^{(1)}\geq\dots\geq z^{(N)}$
    \ENSURE $y = \Pi_{\mathcal{Y}}\left(z\right)$
    \STATE $\Mc_1\leftarrow \emptyset,\Mc_2\leftarrow \{1,\dots,N\},\Mc_3 \leftarrow \emptyset$
\REPEAT 
    \STATE $\rho\leftarrow 2({|\Mc_1|-C+\sum_{n\in \Mc_2} z^n})/{|\Mc_2|}$
    \STATE $y^n\leftarrow\left\{\begin{array}{ll}
1 &  n\in \Mc_1, \\
z^n-\rho/2 & n\in \Mc_2, \\
0 & n\in \Mc_3 \end{array}\right.$
    \STATE $\Sc \leftarrow \left\{n\in\mathcal{N}: y^n<0\right\}$
    \STATE $\Mc_2 \leftarrow \Mc_2\setminus \Sc$, $\Mc_3 \leftarrow \Mc_3\cup \Sc$
\UNTIL{$\Sc=\emptyset$}
        \IF{$y^1 >1$}
        \STATE $\Mc_1\leftarrow \{1\},\Mc_2\leftarrow \{2,\dots,N\},\Mc_3 \leftarrow \emptyset$
        \STATE Repeat 2-7
    \ENDIF \,\,\,\,\,\,\% KKT conditions are satisfied. 
\end{algorithmic}
\end{algorithm}

\subsection{Performance and Relation to Other Policies}\label{sec:FoLreg}

\begin{figure}[t!]
\centering
\subfigure[Fixed stepsize]{\includegraphics[width=1.72in]{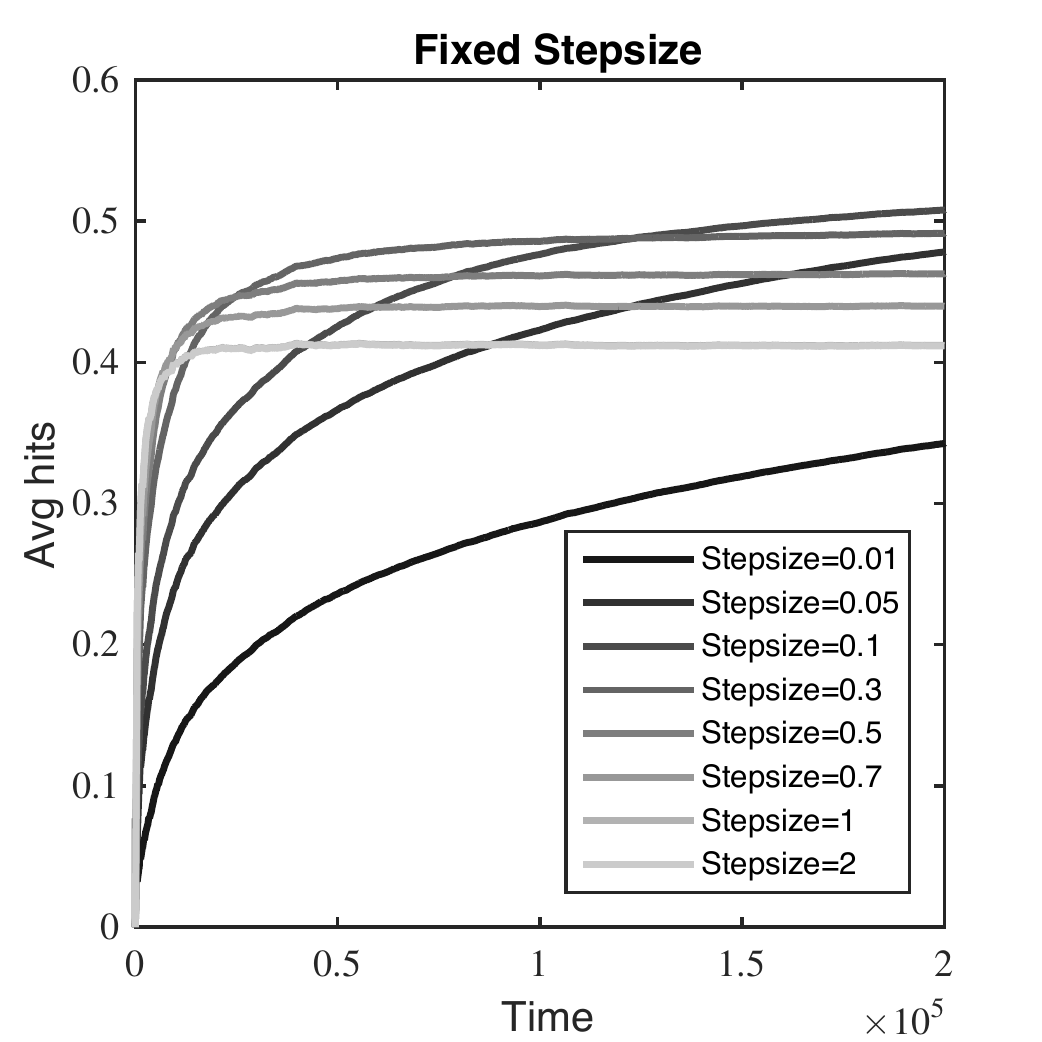}\label{Fig:step1}}
\subfigure[Sorted based on LRU]{\includegraphics[width=1.72in]{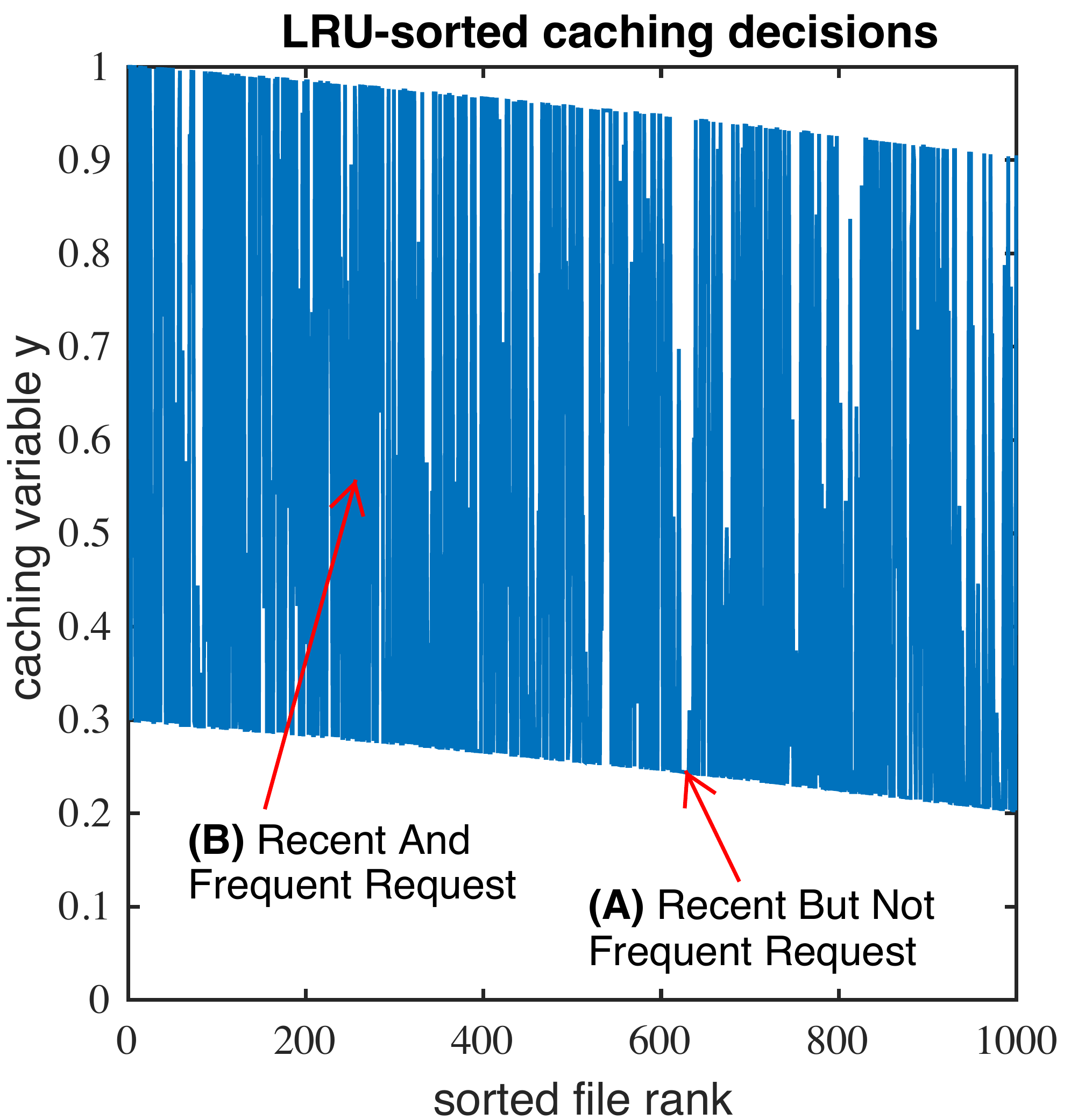}\label{Fig:step4}}
\vspace{-3mm}
\caption{{Single cache with $N\!=\!10^4$, $C\!=\!10^3$.} (a) Smaller stepsizes converge slower, but more accurately. (b) For each file in the LRU cache, we show the respective OGA caching variable.\vspace{-0.1in}}
\label{Fig:stepsize}
\end{figure}

Fig.~\ref{Fig:step1} showcases the hit ratio of OGA for different choices of fixed step sizes, where it can be seen  that larger steps lead to faster but more inaccurate convergence. The horizon-optimal  step is given in Theorem \ref{th:2} as $\eta^*\!=\!w{\sqrt{2C/T}}$, and plugging in $C\!=\!{1000}$, $w\!=\!1$ and $T\!=\!2\cdot10^5$, we obtain $\eta^*\!=\!0.1$; indeed we see that our experiments verify this.

The online gradient descent (similar to OGA) is identical to the {well-known} \emph{Follow-the-Leader} (FtL) policy with a Euclidean regularizer $\frac{1}{2\eta_t}\|y\|$, cf.~\cite{Shalev12}, where FtL chooses in slot $t$ the configuration that maximizes the average utility in slots $1,2,\dots,t-1$. We may observe that FtL {applied here} would cache the files with the highest frequencies, hence it is identical to LFU (when the frequency starts counting from $t\!=\!0$). Hence, OGA can be seen as a regularized version of a utility-LFU policy, where additionally to largest frequencies we smoothen the decisions by a Euclidean regularizer.

\begin{figure*}[!t]
\centering
\subfigure[CDN aggregation (IRM model)]
{\includegraphics[width=1.7in]
{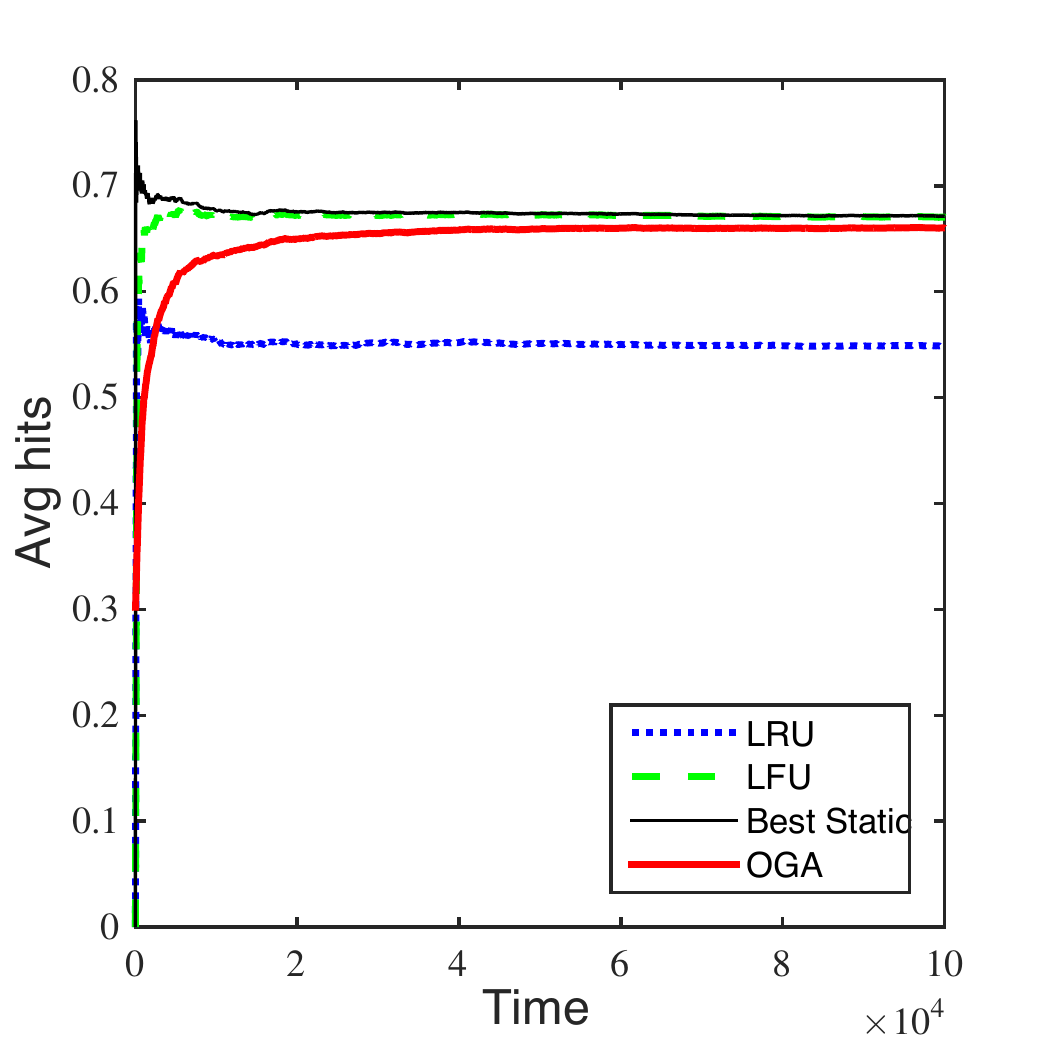}
\label{Fig:Fig1}}
\subfigure[Youtube videos (model \cite{snm})]{\includegraphics[width=1.7in]
{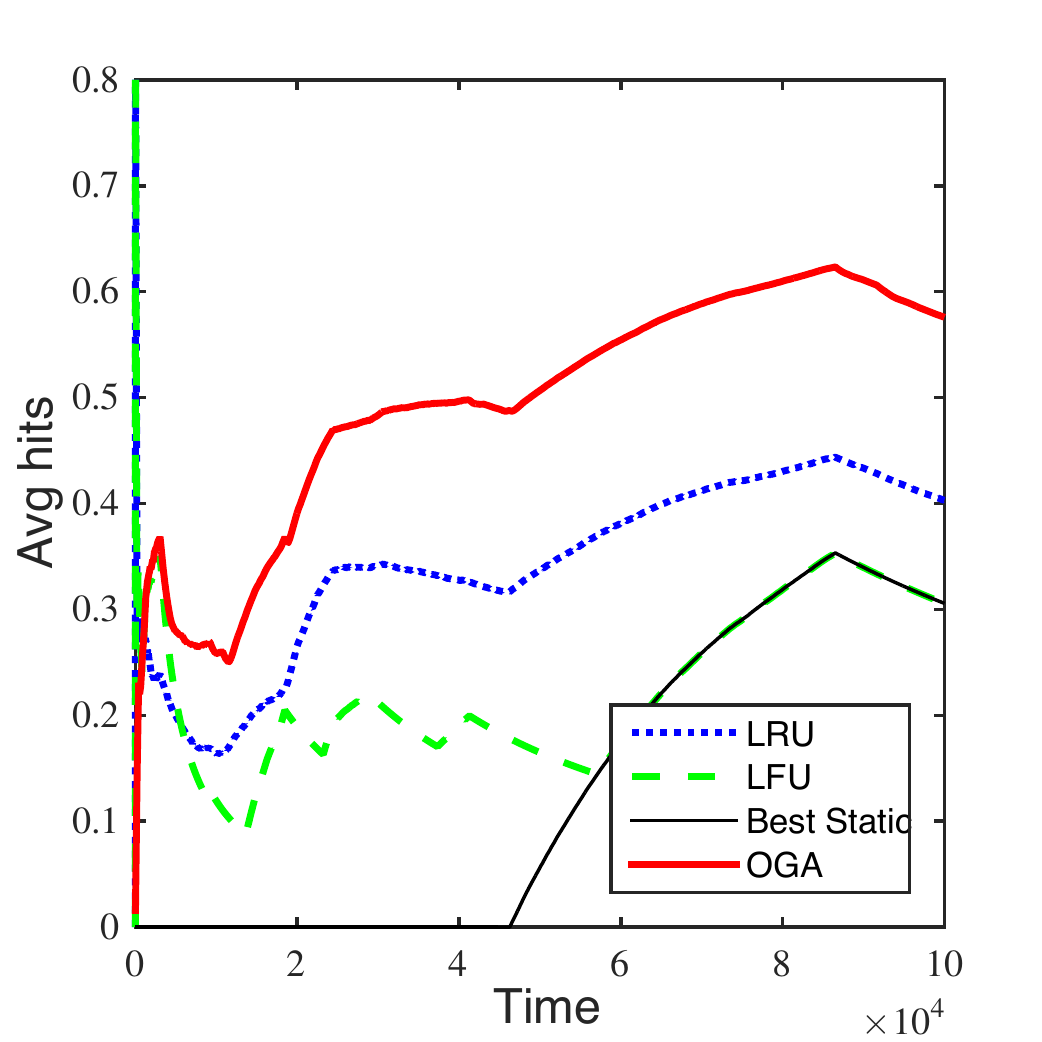}
\label{Fig:Fig2}}
\subfigure[Web browsing (trace \cite{Kurose08})]{\includegraphics[width=1.7in]
{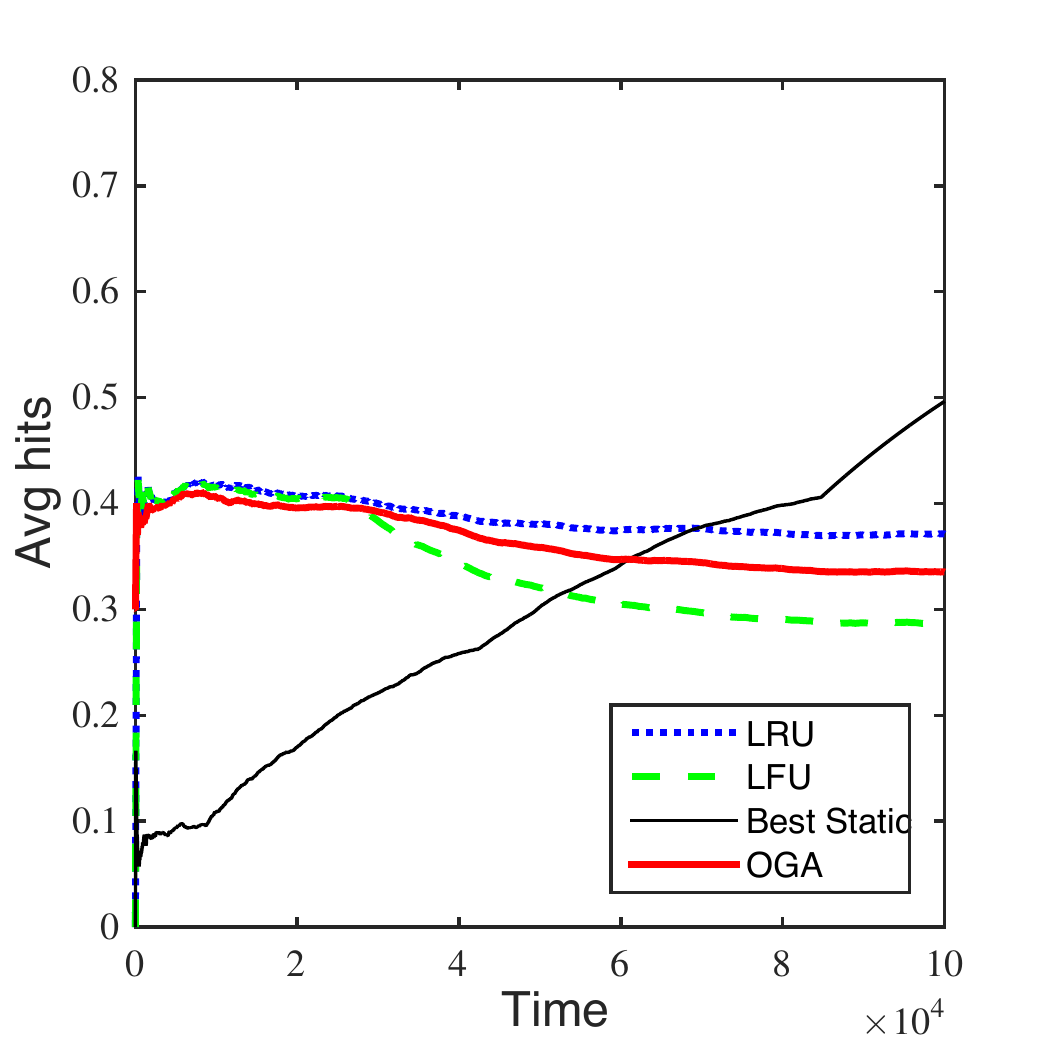}
\label{Fig:Fig3}}
\subfigure[Ephemeral Torrents (model \cite{Elayoubi2015})]{\includegraphics[width=1.7in]
{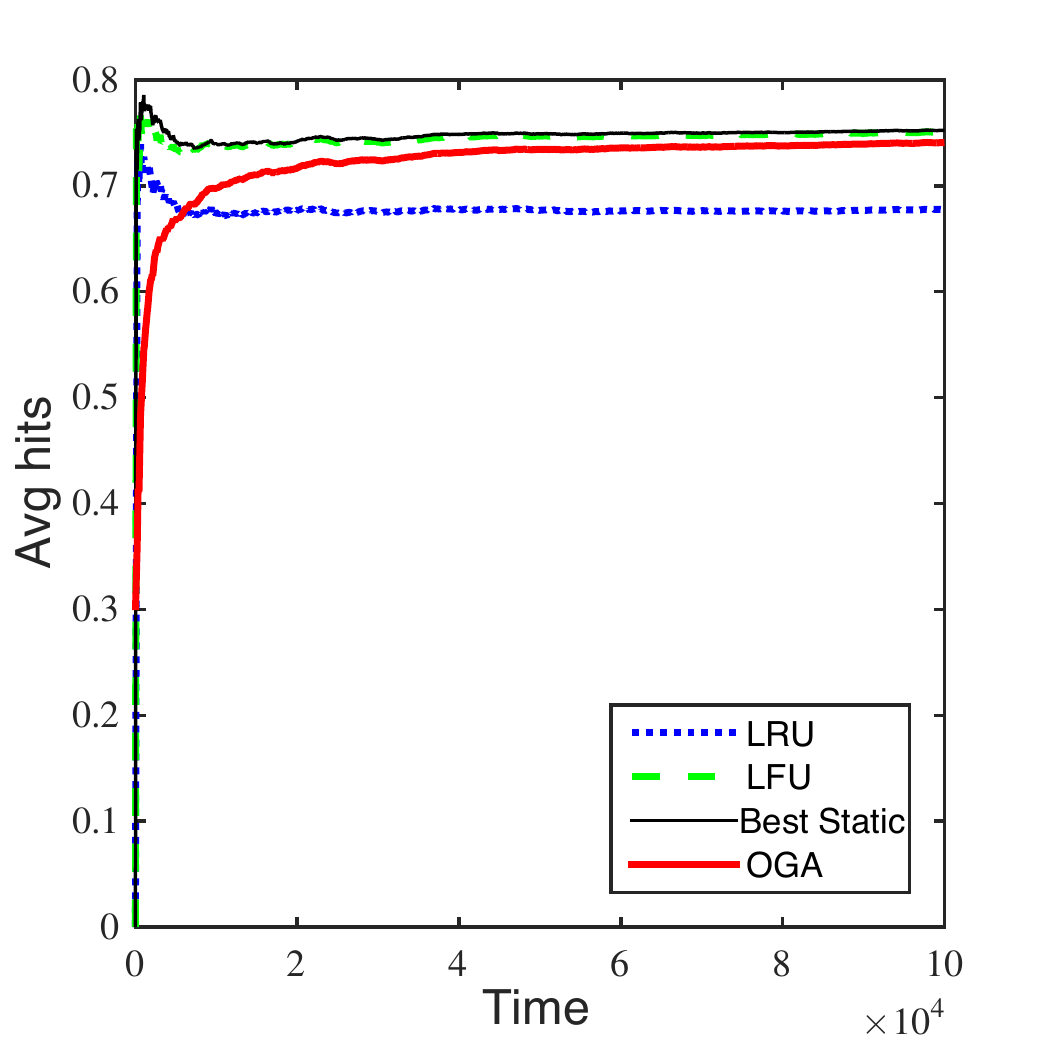}
\label{Fig:Fig4}}
\vspace{-3mm}
\caption{Average hits under different request models \cite{fricker2012impact}; (a) i.i.d. Zipf, (b) Poisson Shot Noise \cite{snm}, (c) web browsing dataset \cite{Kurose08}, (d) random replacement \cite{Elayoubi2015}; Parameters: $\gamma=0.3, T=2\times 10^5, \eta=0.1$.\vspace{-0.0051in}}
\label{Fig:perf2}
\end{figure*}

Furthermore, OGA for $\eta\!=\!1$, $w^n\!=\!w$ bears similarities to LRU, since recent requests enter the cache at the expense of older requests. Since the Euclidean projection drops some chunks from each  file, we expect  least recent requests to drop first in OGA. The difference  is that OGA evicts  files  gradually, chunk by chunk, and not in one shot. Fig.~\ref{Fig:step4} shows the values $y^n(OGA)$ for all files in the  LRU cache ({the $C\!=\!1K$ most recently used}). This reveals that the two policies make strongly correlated decisions, but OGA additionally ``remembers'' which of the {recent requests  are  also infrequent (e.g., see point (A) in Fig. \ref{Fig:step4}), and decreases accordingly the $y^n$ values (as LFU would have done)}.

Finally, in Fig.~\ref{Fig:perf2} we compare the performance of OGA to LRU, LFU, and the best in hindsight static configuration. We perform the comparison for catalogues of $10$K files, with a cache that fits $3$K files, and we use four different request models: \emph{(a)} an i.i.d. Zipf model that represents requests in a CDN aggregation point \cite{fricker2012impact}; \emph{(b)} a Poisson shot noise model that represents ephemeral YouTube video requests \cite{snm}; \emph{(c)} a dataset from \cite{Kurose08} {with actual web browsing requests} at a university campus; and \emph{(d)} a random replacement model from \cite{Elayoubi2015} that represents ephemeral torrent requests. We observe that OGA performance is always close to the best among LFU and LRU. The benefits from the second best policy here is as high as 16$\%$ over LRU and 20$\%$ over LFU.

\vspace{2mm}
\section{Bipartite Online Caching}\label{sec:bipartite}

We extend now our study to a network of caches reachable by the user population via a weighted bipartite graph. Bipartite caching was first used for small cell networks (see the seminal femtocaching model \cite{golrezaei2012femtocaching}), and subsequently  also to model a variety of wired and wireless caching problems where the bipartite graph links represent delay, cost or other QoS metrics when users accessing different caches \cite{paschos-jsac}.

\subsection{Bipartite Caching Model}

Our caching network (CN) comprises a set of user locations ${\cal I}=\{1,2,\dots, I\}$  served by a set of caches ${\cal J}=\{1,2,\dots, J\}$, each with capacity $C_j$, $j\!\in \!{\cal J}$. We use parameters: 
\begin{equation}
\bm{\ell}=\big(\ell_{ij}\in \{0,1\}: i\in\mathcal{I}, j\in\mathcal{J} \big), \notag
\end{equation}
to denote whether cache $j$ is reachable from location $i$, or not ($\ell_{ij}=0$). This includes the general case where a location is connected to multiple caches (e.g., consider base stations with overlapping coverage). We also maintain the origin server as a special node (indexed with $j\!=\!0$), which contains the entire library, and serves the requests for files not found in the caches. The request process is a sequence of vectors with element $x_t^{n,i}\!=\!1$ if a request for file $n$ arrives at $i\in\cal I$ in slot $t$. At each slot $t$ there is only one request $\sum_{n,i}x_t^{n,i}\!=\!1$.

We perform caching using the standard model of \emph{Maximum Distance Separable} (MDS) codes \cite{golrezaei2012femtocaching}, where each stored chunk is a pseudo-random linear combination of original file chunks, and a user requires a fixed number of such chunks to decode the file. Furthermore, we will populate the caches with different random chunks such that, following from the MDS properties, the collected chunks will be linearly independent with high probability (and therefore complement each other for decoding). This results in the following model: the caching decision vector $y_t$ has $N\!\times\!J$ elements, and each element $y_t^{n,j}\in [0,1]$ denotes the amount of random equations of file $n$ stored at $j$. The set of eligible caching vectors is convex: 
\[
\Yc_{\cal J}=\left\{y\in [0,1]^{N\times J} ~\Bigg|~ \sum_{n=1}^{N}y^{n,j}\leq C_j, ~j\in J\right\}.
\]
The caching policy $\sigma$ can be defined as follows:
\begin{equation}
	\sigma: (x_1, x_2, \ldots, x_{t-1}, y_1, y_2, \ldots, y_{t-1})\longrightarrow y_{t}\in \Yc_{\cal J}\,.
\end{equation}

Since each location $i$ might be connected to multiple caches, we use  \emph{routing variables} $z^{n,i,j}_t$ to describe how a request is served at each location. The caching and routing decisions are coupled and constrained: \emph{(i)} a request cannot be routed to an unreachable cache, \emph{(ii)} we cannot route from a cache more data than it has, and \emph{(iii)} each request must be fully routed. Hence, the eligible routing decisions under $y_t$ are: 
\[
{\cal Z}(y_t)=\left\{z\in [0,1]^{N\!\times\! I\!\times \!J} \Bigg|
\begin{array}{c}
\sum_{j\in {\cal J}\cup\{0\}}z^{n,i,j}_t=x^{n,i}_t,~\forall n,i \\
 z^{n,i,j}_t\leq \ell_{ij}y^{n,j}, ~\forall n,i,j\in \mathcal{J}
\end{array}
\right\},
\]
where $z^{n,i,0}_t$ (the routing to origin) is unconstrained, and hence the set ${\cal Z}(y_t)$ is non-empty for all $y_t\!\in\! \mathcal{Y}_{\cal J}$  (i.e. $\sum_{j\in {\cal J}\cup\{0\}}z^{n,i,j}_t\!=\! x^{n,i}_t$ can always be satisfied). Naturally, we assume that $z_t\!\in\! {\cal Z}(y_t)$ is decided after the requests arrive, and hence also after the caching policy. 

Finally, we introduce utility weights $w^{n,i,j}$ to denote the obtained benefit by retrieving a unit fraction of a file from cache $j$ instead of the origin, and trivially $w^{n,i,0}\equiv 0$. Apart from file importance, these weights also model the locality of caches within the geography of user locations--for example a  cache might have higher benefit for certain locations and lower for others. In sum, the total utility accrued in slot $t$ is:
\begin{equation}\label{eq:biput}
f_t(y_t)=\max_{z\in {\cal Z}(y_t)} \sum_{n=1}^N\sum_{i=1}^{I}\sum_{j=1}^{J} w^{n,i,j} x_t^{n,i}z^{n,i,j}_t,
\end{equation}
where the index $t$ is used to remind us that $f_t$ is affected by the adversary's decision $x_t^{n,i}$. The regret of policy $\sigma$ is:
\[
R_T(\sigma)=\max_{P(x_1,\dots,x_T)}\mean{\sum_{t=1}^T f_t(y^*)-\sum_{t=1}^T f_t(y_t(\sigma))},
\]
where $y^*$ is the best static configuration in hindsight (factoring  the associated routing).

Next we  establish the concavity of our objective function $f_t(y)$ for each slot $t$.
{Since} there is only one request at each slot $t$, we can simplify the form of $f_t(y)$. Let $\hat{n}, \hat{i}$ be the file and location where the request in $t$ arrives. Then $f_t( y_t)$ is zero except for $x_{t}^{\hat{n}, \hat{i}}$. Denoting with ${\cal J}^*$ the set of reachable caches from $\hat{i}$, and simplifying the notation, \eqref{eq:biput} reduces to: \vspace{-0.08in}
\begin{align}
f( y )=\max_{z\geq 0 }  \,\,\,&\sum_{j\in {\cal J}^*} w^jz^j \label{eq:opt_zb1}\\
\text{s.t. }& \sum_{j\in \mathcal{J}^*} z^j\leq1 \label{eq:opt_zs}\\
&  z^j\leq \left\{\begin{array}{ll}
y^j &  j\in {\cal J}^* \\
0 & j\notin {\cal J}^*.
\end{array}\right. \label{eq:opt_zc}.
\end{align}

\begin{lemma}\label{lem:convx}
The function $f(y)$ is concave in its domain $\Yc_{\mathcal{J}}$.
\end{lemma}
\emph{Proof:}
Let us consider two feasible caching vectors $y_1,y_2\in\Yc_{\mathcal{J}}$, our goal is to show that:
\[
f(\lambda y_1+(1-\lambda)y_2)\geq \lambda f(y_1)+(1-\lambda) f(y_2),~\forall \lambda \in [0,1].
\]
We begin by denoting with $z_1$ and $z_2$ the routing maximizers of \eqref{eq:opt_zb1} for vectors $y_1,y_2$ respectively. Immediately, it is $f(y_i)=\sum_j w^jz_i^j, ~i=1,2$. Next, consider a candidate vector $y_3=\lambda y_1+(1-\lambda)y_2$ for some $\lambda\in [0,1]$. 
We first show that the routing $z_3=\lambda z_1+(1-\lambda) z_2$ is a feasible routing for $y_3$, i.e., that $z_3\in \mathcal{Z}(y_3)$; by the feasibility of $z_1,z_2$, we have:
\begin{align}
\sum_j z_3^j\!=\!\sum_j (\lambda z_1^j\!+\!(1\!-\!\lambda) z_2^j)\! =\! \lambda \sum_j z_1^j+(1\!-\!\lambda)  \sum_j z_2^j\!=\!1, \nonumber
\end{align}
which proves $z_3$ satisfies \eqref{eq:opt_zs}. Further, for all $j$, it is:
\begin{align*}
z_3^j&=\lambda z_1^j+(1-\lambda) z_2^j \leq \lambda y_1^j+(1-\lambda) y_2^j=y_3^j,
\end{align*}
which proves that $z_3$ also satisfies \eqref{eq:opt_zc}; hence $z_3\in \mathcal{Z}(y_3)$. It follows that $f(y_3)\equiv \max_{z\in \mathcal{Z}(y_3)}\sum_j w^j z^j\geq \sum_j w^j z^j_3$. Combining:
\begin{align}
&f(\lambda y_1+(1-\lambda)y_2)=f(y_3)\geq \sum_j w^j z^j_3= \\
&\lambda \sum_j w^j z^j_1+(1\!-\! \lambda) \sum_j w^j z^j_2\! =\!\lambda f(y_1)+(1\!-\!\lambda) f(y_2).\, \blacksquare \nonumber
\end{align}

\vspace{1mm}
\subsection{Bipartite Subgradient Algorithm (BSA)}

Since $f(y)$ is concave, our plan is to design a universal online bipartite caching policy  by extending OGA. Note, however, that $f(y)$ is not necessarily differentiable everywhere (hence the gradient might not exist at $y_t$), and hence we need to find a subgradient direction at each slot, as explained next.

Consider the partial Lagrangian of \eqref{eq:opt_zb1}: 
\begin{align*}
	L(y,z,\alpha,\beta)&=\sum_jw^jz^j+\alpha\big(\sum_j z^j-1\big)+\sum_{j}\beta^j(z^j-y^j),
\end{align*}
and define  the  function $\Lambda (y,\beta)\!=\!L(y,z^*,\alpha^*,\beta)\!\equiv\! \min_{\alpha\geq 0}\max_{z\geq 0}	L(y,z,\alpha,\beta)$.
From strong duality we obtain:
\begin{equation}\label{eq:strdual}
f(y)=\min_{\beta \geq 0}\Lambda (y,\beta).
\end{equation}

\begin{lemma}[Supergradient]\label{lem:subgradient}
Let $\beta^*(y)\triangleq \arg\min_{\beta\geq 0}\Lambda (y,\beta)$ be the vector of optimal multipliers of  \eqref{eq:opt_zc}. Define:
\[
g^j=\left\{\begin{array}{ll}
\beta^{j,*}(y) & j\in {\cal J}^* \\
0 & j\notin {\cal J}^*,
\end{array}\right.
\]
The vector $g$ is a supergradient of $f$ at $y$, i.e., it holds $f(y)\geq f(y')-g^T(y'-y),~\forall y'\in \Yc_{\cal J}$.
\end{lemma}
\begin{IEEEproof}
We have
\begin{align*}
f(y)&\stackrel{\eqref{eq:strdual}}{=}\min_{\beta \geq 0}\Lambda (y,\beta)\equiv\Lambda (y,\beta^*(y))\\
& \stackrel{(a)}{=}\Lambda (y',\beta^*(y))-\beta^*(y)^T(y'-y)\\
& \stackrel{(b)}{\geq} \Lambda (y',\beta^*(y'))-\beta^*(y)^T(y'-y) \\
&=f(y')-\beta^*(y)^T(y'-y)
\end{align*}
where $(a)$ is obtained directly by the form of the Lagrangian $L(\cdot)$ where only one term depends on $y$, and in (b) we have used that   $y'$ minimizes  $\Lambda(;,\beta^*(y'))$. 
\end{IEEEproof}

{Now that we have found a method to calculate the supergradient, we can extend OGA as follows:}

\begin{definition}[BSA]
The {Bipartite Subgradient Algorithm} caching policy adjusts the decisions with  the supergradient:
\[
y_{t+1}=\Pi_{\mathcal{Y}_{\cal J}}\left(y_t+\eta_t g_t\right),
\]
where $\eta_t$ is the step, $g_t$ is given in Lemma~\ref{lem:subgradient}, and $\Pi_{\mathcal{Y}_{\cal J}}\left(z\right)\triangleq \argmin_{y\in\Yc_{\cal J}}\|z-y \|$ 
is the Euclidean projection of the argument vector $z$ onto $\mathcal{Y}_{\cal J}$, performed with Algorithm 1.
\end{definition}

\vspace{0.06in}
\begin{tcolorbox}[boxrule=0.7pt,arc=0.6em] 
\begin{theorem}[Regret of BSA]\label{th:3}
Let $C_j=C<N/2$. The regret of \textup{BSA} satisfies:\vspace{-0.1in}
\[
R_T(\textup{BSA})\leq 
w^{(1)}\sqrt{2\textup{deg} J CT}.
\]
\end{theorem}
\end{tcolorbox}
\vspace{0.06in}
\begin{IEEEproof}
Replacing the gradient with the supergradient and repeating the steps of Theorem \ref{th:2} proof, we arrive at: 
\begin{equation}\label{eq:telesub}
\sum_{t=1}^Tg^T_t(y^*-y_t)\leq \frac{\textit{diam}(\Yc_{\cal J})^2}{2\eta} +\frac{\eta TL^2}2,
\end{equation}
where  it holds $f_t(y_t)\geq f_t(y)+ g^T_t(y_t-y)$, $\forall y\in\mathcal{Y}_{\cal J}$. Plugging these in the BSA regret expression we get:
\begin{align*}
R_T(\textup{BSA})&=\sum_{t=1}^T(f_t(y^*)-f_t(y_t)) =\sum_{t=1}^Tg^T_t(y_t-y^*)\\
&\stackrel{\eqref{eq:telesub}}{\leq} \frac{\textit{diam}(\Yc_{\cal J})^2}{2\eta} +\frac{\eta TL^2}2,
\end{align*}
where, $\textit{diam}(\mathcal{Y}_{\cal J})=\sqrt{\sum_{j\in \mathcal{J}}C_j}=\sqrt{2CJ}$ (for $C_j=C<N/2$) and $L$ is  an upper bound on the supergradient norm:
\[
 \|g\|\leq \sqrt{\sum_{j:\ell_{ij}=1}(w^{(1)}-w^{(J)})^2}=w^{(1)}\sqrt{\text{deg}}\triangleq L,
\]
where $\text{deg}$ is the maximum right degree and $w^{(1)}$ the maximum utility, and we used the optimal constant stepsize.
\end{IEEEproof}

The functionality of BSA is similar to OGA. In a slot where file $n'$ has been requested from location $i'$, BSA will increase variables $y^{n',j}$, where $j$ is reachable by $i'$, and then each cache $j$ will perform a local  projection step, decreasing all other variables $y^{n,j}>0$ until the local cache constraint is satisfied. The difference from OGA lies only in how we compute  the $y^{n',j}$ increase. This now depends on the corresponding subgradient element, which is proportional to the hypothetical utility gain if we would increase $y^{n',j}$ by $\delta y$ and we would launch another request for $n'$ from the same location $i'$. 

We have performed extensive simulations of BSA and selected to present a representative scenario in Fig.~\ref{Fig:bsa1}. The bipartite graph is composed of 3 caches with $C=10$, and 4 user locations. The caches have utilities (1,2,100), and a good caching policy should place popular content on cache 3. The network is fed with stationary Zipf requests from a catalogue of 100 files, and each request arrives at a user location uniformly at random. We compare  BSA to the best static policy in hindsight (we may verify BSA has no regret), and literature heuristics mLRU from \cite{giovanidis-mLRU} and $q$-LRU with the lazy rule from \cite{leonardi-implicit} (for $q=1$). Over time, BSA learns what files are popular and increases their cache allocation at the high utility cache, while adjusting accordingly the less popular files in other caches. {Doing so, BSA outperforms lazy-LRU, which was the best heuristic in our simulations,  by $45.8\%$.

\begin{figure}[!t]
	\centering
	\subfigure[Bipartite scenario] 
{\includegraphics[width=1.5in]{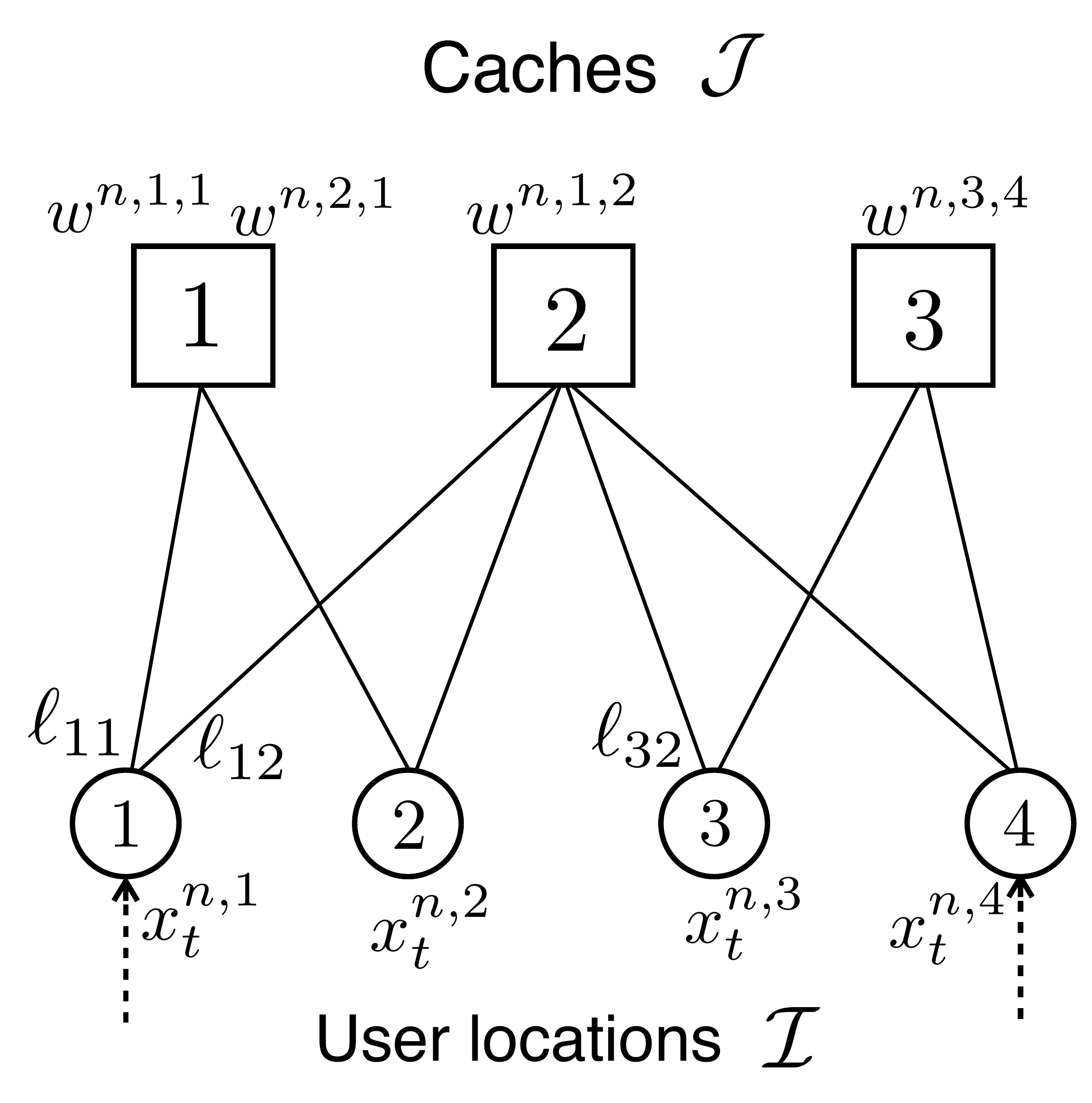}\label{Fig:bsa1}}\,\,\,\,\,\,
\subfigure[Simulation results]{\includegraphics[width=1.7in]{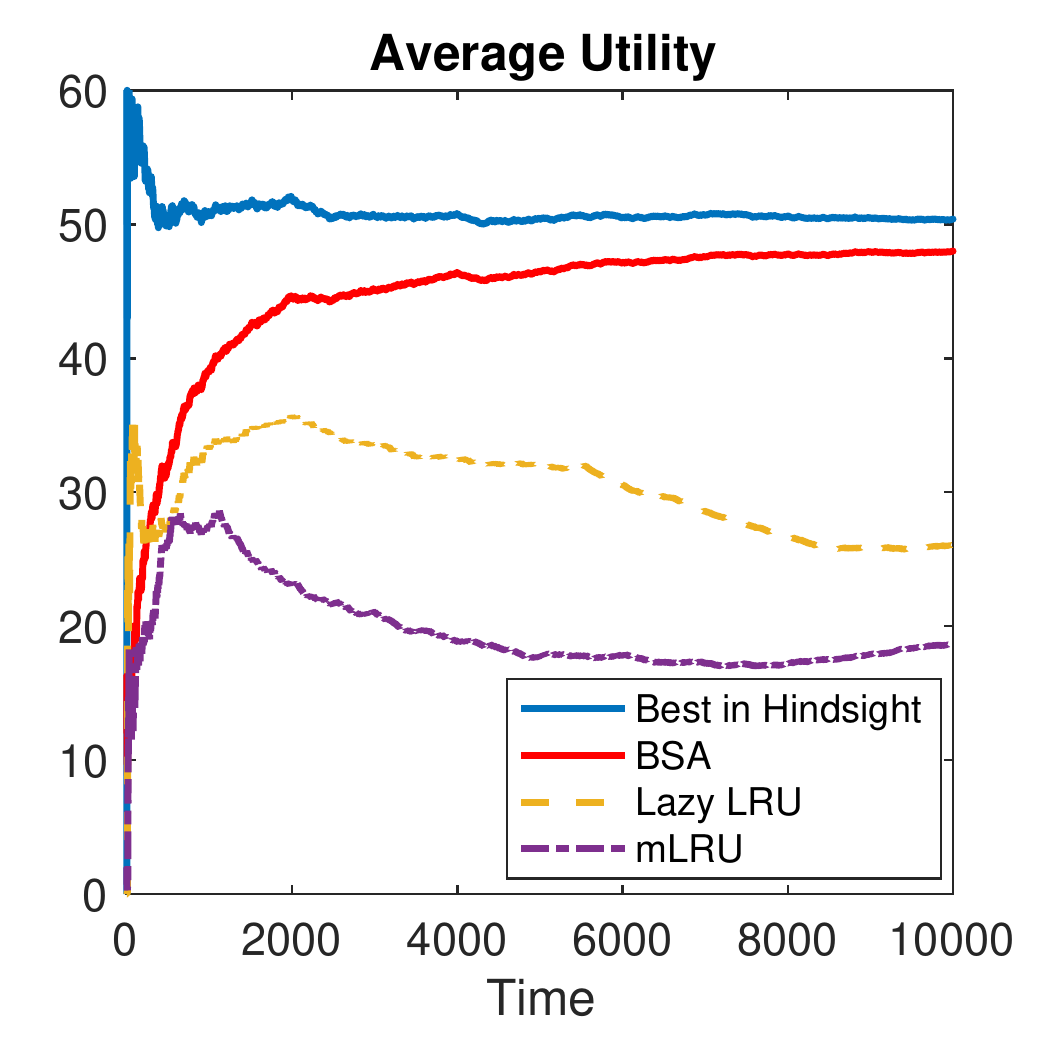}
\label{Fig:bsa2}}
	\caption{(a) A bipartite caching example with 3 caches and 4 user locations.  (b) Average utility  of different caching policies.
	\vspace{-0.2in}
	}
	\label{Fig:bipartite}
\end{figure}

\vspace{1mm}
\section{Conclusions}\label{sec:conclusions}
 
We established a connection between caching policies such as LRU, LFU and their variants, and the framework of OLO. Our analysis sheds light from a different angle to these established (but based on intuition) policies, and hence allows us not only to characterize their performance but also to re-design them. To this end, we proposed the online gradient caching algorithm and proved that it has universally optimal performance; and followed a similar OLO-based approach for bipartite caching networks where routing decisions are also considered. We believe our work opens a very exciting area, paving the road for the principled analysis and design of dynamic caching policies for single or networks of caches.

%
%
%
\bibliography{mybib_v13}
\bibliographystyle{IEEEtran}
\end{document}